\documentclass{article}
\usepackage{ijcai25}

\usepackage{times}
\usepackage{soul}
\usepackage{url}
\usepackage[hidelinks]{hyperref}
\usepackage[utf8]{inputenc}
\usepackage[small]{caption}
\usepackage{graphicx}
\usepackage{amsmath}
\usepackage{amsthm}
\usepackage{booktabs}
\usepackage{epstopdf}
\usepackage{algorithm}
\usepackage{algorithmic}
\usepackage[switch]{lineno}
\usepackage{amsmath,amssymb,amsfonts}
\usepackage{xcolor}

\title{Mentor-Telemachus Bond: Transferring Knowledge in Semantic Communication via Contrastive Learning\\}

\author{
Zhiyuan Xi \and
Kun Zhu \and
Yuanyuan Xu \and
Tong Zhang
\emails
xizhiyuan@nuaa.edu.cn,
zhukun@nuaa.edu.cn,
yuanyuan\_xu@hhu.edu.cn,
zhangt@nuaa.edu.cn
}

\begin{document}

\maketitle

\begin{abstract}
Encoder, decoder and knowledge base are three major components for semantic communication. Recent advances have achieved significant progress in the encoder-decoder design. However, there remains a considerable gap in the construction and utilization of knowledge base, which plays important roles in establishing consensus among communication participants through knowledge transferring and sharing. Current knowledge base designs typically involve complex structures, which lead to significant computational overheads and heavy reliance on manually annotated datasets, making it difficult to adapt to existing encoder-decoder models. Hence, without knowledge transferring and sharing within the network results in poor generalization of encoder-decoder. This necessitates model training for specific tasks and datasets, significantly limiting the scalability of semantic communication systems to larger networks. To address these challenges, we propose an innovative Contrastive Representations Learning based Semantic Communication Framework (CRLSC). In CRLSC, the server-side pre-trained large model utilizes large-scale public datasets to construct shared knowledge base. Local-side encoders in terminal devices conduct training guided by shared knowledge base. These trained encoders can then build private knowledge bases from private datasets and fine-tune decoders for specific tasks. This simple and effective approach can facilitate the knowledge transferring across large-scale heterogeneous networks. 
\end{abstract}

\section{Introduction}
The information theory proposed by Shannon and Weaver~\cite{Shannon,Weaver} forms the cornerstone of modern communication technologies. Their theory categorizes communication into three levels: syntax, semantics, and pragmatics. These levels focus on the accurate transmission of symbols, the precise conveyance of the meanings or intentions behind the transmitted information, and the effective utilization of that information, respectively.

\begin{figure*}[htbp]
    \centering
    \noindent\includegraphics[width=0.7\textwidth]{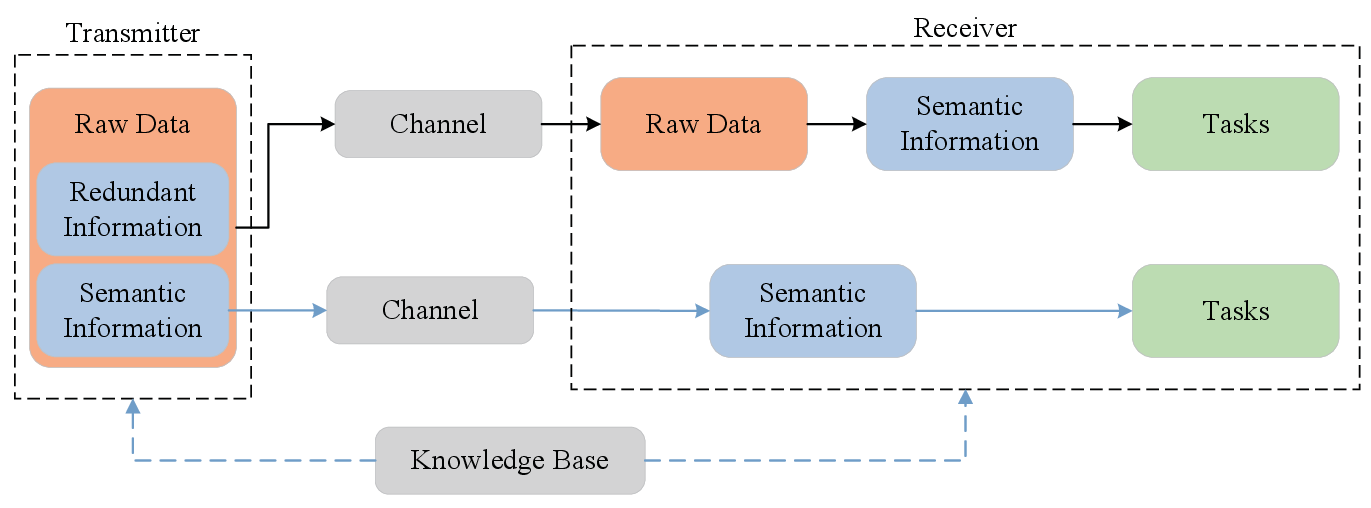}
    \caption{\textbf{The comparison between conventional communication and semantic communication}. The upper branch represents conventional communication, while the lower branch is semantic communication filtering out redundant information from the raw data.}
    \label{fig_1}
\end{figure*}

As shown in Fig.~\ref{fig_1}, conventional communication focuses on the accurate transmission of symbols. However, it fails to extract and convey the key semantics that the receiver requiring, leading to the transmission of a large amount of redundant information and imposing additional burden on the communication system. Besides, due to differences in background knowledge, the transmitter and receiver may have varying cognitions, which can lead to the receiver misinterpreting the information from the transmitter. This misinterpretation could hinder the accurate completion of various tasks. Therefore, it is essential for communication technology to evolve to the next level---semantic communication, to meet the complex demands of modern AI-driven applications.

Naturally, semantic communication has attracted considerable interests from academia and industry.  To address the limitations of conventional communication, the semantic communication framework~\cite{survey_1,survey_2} designs three core components: an encoder, a decoder, and a knowledge base. The encoder extracts semantic information from the raw data, capturing the underlying meaning. The decoder then uses the received semantic information to execute various tasks, with information reconstruction being the most common task. Utilizing encoder and decoder, the transmitter and receiver can bypass the transmission of raw data, focusing solely on the transmission of key semantics. This approach significantly reduces the communication overheads in conventional communication. Currently, Researchers have developed a variety of semantic communication frameworks for encoder and decoder, including those based on Transformers~\cite{transformer}, Diffusion models~\cite{diffusion} and VQ-VAE~\cite{vq-vae}.

The knowledge base can leverage stored public data and historical task information to establish consensus among communication participants through knowledge transferring and sharing. This alleviates misunderstandings among communication participants and drives the expansion of semantic communication systems to larger-scale networks. However, there still exists several challenges: \textbf{(1)} Most work considers the design of encoder-decoder and knowledge base independently. Driven by the rapid development of deep neural networks, research tends to focus on encoder and decoder design while neglecting the role of knowledge base; \textbf{(2)} Research on knowledge base often involves complex structures and computations, such as knowledge graphs and entity relationship extraction. This results in significant computational overheads and a heavy dependence on manually annotated datasets, making it difficult to adapt to existing encoder/decoder models; \textbf{(3)} The lack of knowledge transferring and sharing means that encoders and decoders can only be trained end-to-end for specific tasks and datasets, which results in poor generalization capability.

In this paper, we gain insight into the current semantic communication systems and mainstream deep learning technologies. Depending on the substantial breakthrough for the AI community in pre-trained models, such as BERT~\cite{bert} in natural language processing, ViT~\cite{vit} in image processing and CLIP~\cite{clip} in multimodal processing, we could deploy these pre-trained models on high-performance servers. These pre-trained models can transform large-scale datasets into representation vectors, which are then stored in a vector database to construct a shared knowledge base.

The shared knowledge base encapsulates the data comprehension of the pre-trained model . By leveraging shared knowledge base, we can facilitate the training of local encoder, establishing a novel distributed training paradigm. Reflecting on the process of human learning, students often seek help from mentors when encountering new concepts, gaining insights from their knowledgeable mentors. For example, a student who has never seen a zebra can enhance their understanding by learning about horses from a mentor.

Contrastive learning employs data augmentation across two branches to generate distinct views of the same sample. These views are referred to as positive pairs, while views from different samples are considered negative pairs. During training, the model is optimized to minimize the distance between positive pairs and maximize the distance between negative pairs. This self-supervised approach enables the model to learn distinctive features for each instance, thereby enhancing its generalization capabilities. Inspired by this concept, we utilize the shared knowledge base to provide one of the sample views for one branch, similar to a mentor offering guidance to a student, thus aiding the training of the local encoder. Once the local encoder has completed training, task-specific decoders can be incorporated and fine-tuned, forming a semantic communication system that is adaptable to various tasks. Moreover, by leveraging the local encoder, we can construct a private knowledge base and facilitate knowledge transfer between terminal devices through the same contrastive learning process.

To the best of our knowledge, this is the first study that provides a comprehensive exploration of the knowledge base construction and practical implementation using a straightforward and easily implementable method. The main contributions of this paper can be summarized as follows:

\begin{itemize}
    \item We develop a simple yet effective approach for constructing a knowledge base in semantic communication. This involves using a large, server-side pre-trained model to convert a large-scale dataset into representation vectors, which are then stored in a vector database to construct the shared knowledge base.
    \item We adopts an innovative contrastive learning methods to transfer knowledge from shared knowledge base to local models, which significantly reduces the computational and storage requirements for training models locally.
    \item The approach we proposed for knowledge base construction and knowledge transferring exhibits excellent versatility. It could be seamlessly integrated into existing encoder-decoder research and scale the semantic communication system to the larger-scale networks.
\end{itemize}

\section{Related Work}
\subsection{Semantic Communication System}
\subsubsection{Encoders and Decoders}
Currently, the majority of research is focused on the encoders and decoders. Transformer is adopted in~\cite{sc_enc_dec_1} and~\cite{sc_enc_dec_2} to replace the classical neural network architectures. With the self-attention mechanism, a diverse set of semantic communication problems could be well addressed. Besides, a squeeze-and-excitation network is employed in~\cite{sc_enc_dec_3} to transmit speech signals. To tackle the challenges of speech recognition and synthesis in communication system, CNN and RNN are employed in~\cite{sc_enc_dec_4} for semantic encoding, and dense layers for channel encoding and decoding. Considering multi-user scenarios,~\cite{sc_enc_dec_5} designs a swin transformer-based dynamic semantic communication framework. To combat the semantic noise, a masked VQ-VAE model is adopted in~\cite{sc_enc_dec_6}, which significantly improves the system robustness.

\subsubsection{Knowledge Base}
Despite being crucial in semantic communication, research on knowledge base is still scarce. A cloud-edge-device collaborative knowledge base is proposed in \cite{sc_kb_1}.~\cite{sc_kb_2} designs a shared knowledge base to extract keywords from raw data. These keywords are encoded by auto-encoder and decoded by auto-decoder, and then raw data are recovered depending on shared knowledge base and received keywords. Similarly, a textual knowledge base is proposed in~\cite{sc_kb_3} for providing residual information related to the transmitted messages. To meet the high reliability and low latency demands in vehicular network scenarios, comprehensive analyses of knowledge matching-based queuing latency for semantic data packets have been provided in~\cite{sc_kb_4}. Additionally, a novel application of knowledge bases has been introduced to prevent the leakage of sensitive information in~\cite{sc_kb_5}. These studies have initially explored the knowledge base, however, the knowledge bases in these studies often involve complex structures, such as knowledge graphs, which require additional manual data annotation and cleaning. These issues make it difficult for knowledge base systems to be implemented and replicated, thereby hindering further research. Furthermore, the sources of knowledge are relatively narrow, and there is a lack of a unified approach for extracting knowledge from multimodal data.

\subsection{Contrastive Learning}
The basic idea of contrastive learning involves creating pairs or sets of related and unrelated data samples. For each "positive" pair (samples that are similar or related), the model learns to pull the representations together. For each "negative" pair (samples that are dissimilar or unrelated), the model learns to push their representations apart. To simplify the design of contrastive learning,~\cite{cl_1} discards the memory bank, proposing a simplified contrastive learning framework. Furthermore, empirical evidence demonstrates that data augmentation is an effective method for learning robust representations. Building on these pioneering discoveries, ~\cite{cl_2},~\cite{cl_3},~\cite{cl_4} and~\cite{cl_5} introduce further improvements and advancements. It is noteworthy that~\cite{cl_6} introduces the concept of a support set, through which the training of positive and negative samples is accomplished by matching approximate representations within this set, yielding quite impressive results. Furthermore, the CLIP model proposed in~\cite{clip} pioneered the field of multimodal contrastive learning and has become the cornerstone of various large language models. It is trained on a wide variety of images paired with their corresponding textual descriptions from the internet. CLIP employs contrastive learning to project text and image representations into a shared space. With minimal fine-tuning of task-specific modules, these representations can effectively be used for various downstream tasks, such as image classification and object detection.

\begin{figure*}[htbp]
    \centering
    \noindent\includegraphics[width=6in]{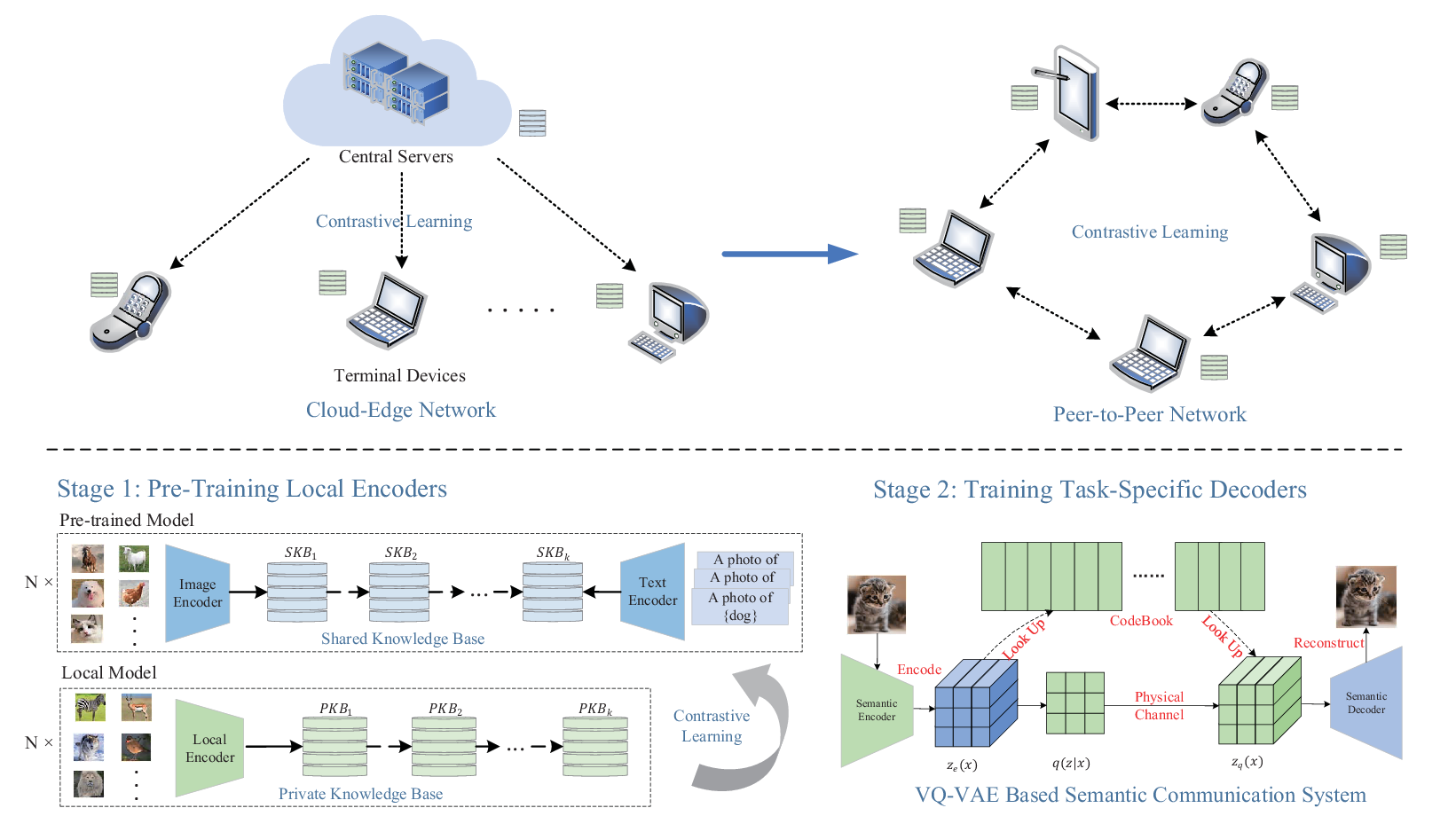}
    \caption{Contrastive Representations Learning based Semantic Communication Framework}
    \label{fig_3}
    \vspace{-5pt}
\end{figure*}

\section{Contrastive Representations Learning based Semantic Communication Framework}
Semantic communication involves the construction and transmission of multimodal data, with representation vectors serving as the unified form of these data. Multimodal data can be converted into representation vectors by representation learning models. These vectors can then be effectively utilized in various downstream tasks. Vector quantization provides the essential support for efficient storage of these representation vectors. 

Based on the above considerations, as shown in Fig.~\ref{fig_3}, we design an innovative Contrastive Representations Learning based Semantic Communication Framework (CRLSC). CRLSC enables terminal devices to perform pre-training of local models in cloud-edge networks under the guidance of a server-side shared knowledge base. This paradigm can be further extended to peer-to-peer networks, where terminal devices facilitate knowledge transferring through private knowledge bases. Regardless of the network type, the training of local models is divided into two primary stages: \textit{Stage 1} encompasses the pre-training of local encoder, while \textit{Stage 2} involves the training of task-specific decoders based on these pre-trained encoders, thereby constructing a semantic communication system. Next, we provide a detailed description of each stage.

\subsection{Stage 1: Pre-Training Local Encoder}

\subsubsection{Constructing Shared Knowledge Base}
Before introducing the knowledge base, it is essential to demonstrate the concept of knowledge. According to~\cite{concept_knowledge}, knowledge can be categorized into parametric and non-parametric types. Parametric knowledge refers to model parameters acquired through data training, encapsulating the model comprehension of the data. Conversely, non-parametric knowledge pertains to external knowledge, such as the information stored in a vector database, which lies outside the scope of the training data. In the CRLSC, the definition of the knowledge can be summarized as follows:

\begin{itemize}
    \item \textbf{Parametric knowledge}: We consider the pre-trained model on the server side and the semantic encoder on the local side as parametric knowledge, denoted by $S_e$ and $L_e$ respectively. These models are trained on large amounts of data, and their parameters encapsulate an understanding of external information.
    \item \textbf{Non-parametric knowledge}: The shared knowledge base is denoted by $\mathit{SKB}$, while the private knowledge base is $\mathit{PKB}$. Both of them are non-parametric knowledge. It is worth noting that the number of servers and clients mentioned in the CRLSC is not limited.
\end{itemize}

On the server side, there are some public datasets denoted by $\mathcal{D}=\left\{d_1, d_2, \dots, d_k \right\}$, and $d_k$ represents a type of unimodal dataset. Datasets are encoded into feature vectors by $S_e$ and stored in vector knowledge bases, serving as the shared knowledge bases denoted by $\mathcal{S}=\left\{\textit{SKB}_1, \textit{SKB}_2, \dots, \textit{SKB}_k \right\}$. Similarly, the local private knowledge base can be constructed in this manner by $L_e$, represented by  $\mathcal{P}=\left\{\textit{PKB}_1, \textit{PKB}_2, \dots, \textit{PKB}_k \right\}$. 

In the CRLSC, we adopt the CLIP~\cite{clip} as the pre-trained large model at the server side. CLIP is equipped with two distinct encoders: an image encoder and a text encoder. We feed a large-scale dataset, consisting of images from multiple categories, into the image encoder.  Moreover, following the approach in CLIP, we construct category information into sentences format as \textit{'a photo of \{label\}'} and input them into the text encoder. Subsequently, these data are transformed into representation vectors by CLIP. Then, a Product Quantization (PQ)~\cite{vq_1} based vector database is employed to quantize and store these vectors efficiently. Suppose a feature vector $\mathbb{V}^d$, then we divide $\mathbb{V}^d$ into $m$ subvectors $\mathbb{V}^{d^{*}}$, where $d^{*} = d / m$. Subvectors in each subspace are clustered separately, and the number of cluster centers in each subspace is denoted by $k^*$. Hence, the storage complexity could be calculated as $m \times d^* \times k^*$. Meanwhile, we can easily calculate that the total number of vectors that can be approximately represented is $k=(k^*)^m$.

At this point, the construction of the shared knowledge base is complete. This process allows us to transform the parametric knowledge of the pre-trained large model into non-parametric knowledge.

\subsubsection{Data Augmentation}
On the local side, there is a small model whose number of parameters is significantly less than that of the server-side pre-trained model. Following the SimCLR~\cite{cl_1}, we adopt the composition of the data augmentation: random resized crop, random horizontal flip, color jitter, gaussian blur and random gray scale. Through such approach, local model could learn generalized representations. 

\subsubsection{Contrastive Learning}
\begin{figure}[htbp]
    \centering
    \noindent\includegraphics[width=3.5in]{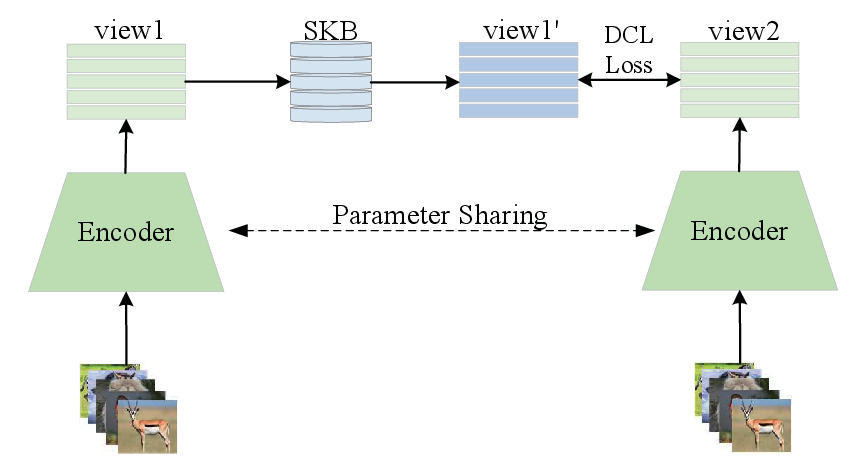}
    \caption{Main Learning Process}
    \label{fig_4}
\end{figure}

The local encoder has two branches with shared parameters, as shown in Fig.~\ref{fig_4}. The left branch encodes augmented data into the representation vectors $q \in \mathbb{R}^{B \times 1 \times d}$, which are used to retrieving the $\textit{SKB}$. The right branch encodes augmented data into the representation vectors $p \in \mathbb{R}^{B \times 1 \times d}$, which do not require retrieving the $\textit{SKB}$. According to the rules of the PQ retrieval algorithm, the vector $q$ is divided into $m$ sub-vectors, each of which calculates distances from the centroids in PQ. By summing these distances, we can obtain the top-n nearest approximate vector representations. Besides, to broaden the search scope and enhance the diversity of retrieval results, we add Gaussian noise $\mathcal{N}(\mu, \sigma^2)$ into $q$, which could be written as $q = \mathcal{N}(\mu, \sigma^2) + q$. 

Then, we denote the retrieval result as vector $k = v \in \mathbb{R}^{B \times n \times d}$ and cross attention is adopted to compute the attention score between $q$ and $k$, which could be written as:
\begin{equation}
\begin{split}
score = \frac{q \times k ^ T}{\sqrt{d}},
\end{split}
\end{equation}
Ultimately, we can merge $q$ with the retrieval results $v$ according to the following formula:
\begin{equation}
\begin{split}
q^* = score \times v.
\end{split}
\end{equation}

Thus, the local side can use the enhanced vector $q^*$, enriched with $\mathit{SKB}$ knowledge, to conduct contrastive learning training.

In contrastive learning, the encoder encodes two views of the data. The enhanced $q^*$ obtained previously can serve as the representation vector for one view, while the representation vector for the other view is $p$, which directly outputed by the encoder.

We adopts the Decoupled Contrastive Learning Loss (DCLloss)~\cite{DCLLoss} to assist the local encoder in distinguishing between positive and negative samples (where positive samples refer to two views of the same data, and negative samples refer to the views of different data in the mini-batch), which can be written as follows:
\begin{equation}
\begin{split}
DCLloss=-log\frac{e^{\langle{z}, {z_{+}}\rangle /\tau}}{\sum\limits_{i=1}^{K}e^{\langle{z}, {z_{-}^{(i)}}\rangle /\tau}},
\end{split}
\end{equation}
where $\langle{z}, {z_{+}}\rangle$ denotes the dot-product of positive pairs, $\langle{z}, {z_{-}^{(i)}}\rangle$ denotes dot-product of negative pairs and $\tau$ is temperature parameter. Compared to the commonly used InfoNCE loss in contrastive learning, DCLloss requires lower batch sizes and fewer epochs for training. This makes it suitable for training on devices with limited computational and storage capabilities.

Through multiple training iterations, we aim to minimize the distance between positive samples and maximize the distance between negative samples.

As observed, in the \textit{stage 1}, the form of knowledge transitions from the parametric knowledge of the pretrained model to the non-parametric knowledge stored in the vector database. Furthermore, under the supervision of non-parametric knowledge, the local encoder is able to learn parametric knowledge, which can then be further transform the local datasets into representation vectors stored in $\textit{PKB}$ as non-parametric knowledge. We can further leverage $\textit{PKB}$ to guide other devices in the network for contrastive learning, facilitating the dissemination of knowledge across the entire network.

This transformation of knowledge form brings a certain degree of decoupling to the entire system. Non-parametric knowledge can be transferred to devices with lower computational power for storage, while devices with substantial computational resources can focus solely on learning new data and updating parametric knowledge. Consequently, the network architecture of semantic communication systems can become more flexible. Especially in large-scale distributed systems, our proposed method facilitates the exchange of knowledge between devices with significant differences in computing and storage capabilities, thereby alleviating the issue of data silos to a certain extent.

Although the method we proposed bears some similarities to knowledge distillation, in that it involves transferring knowledge from a pretrained large model to a smaller model, there are fundamental differences. In knowledge distillation, the smaller model directly mimics the larger model, and both typically operate on the same dataset, with the larger model continuously generating soft labels to provide supervisory signals. However, our method considers not only the transfer of knowledge from the large model, represented as $q^*$, but also the data understanding of the small model, represented as $p$. Moreover, the datasets for the large and small models can be different.

\subsection{Stage 2: Training Task-Specific Decoders}
Typically, semantic communication can be viewed as an information reconstruction task, where the encoder is used to extract key features (semantic information) from the raw data, and the decoder utilizes these features to reconstruct the original information. 

Since this paper primarily focuses on the design and construction of knowledge base, for simplicity, we adopt the VQ-VAE to construct the semantic communication system, following the approach of~\cite{sc_enc_dec_6}. In the VQ-VAE based semantic communication system, a raw image is initially encoded by the encoder into $z_e(x) \in \mathbb{R}^{k \times d}$. The encoded representation $z_e(x)$ is then matched with vectors $e$ stored in a CodeBook by calculating the Euclidean distance.  The index of the vector with the smallest distance is used to discretize $z_e(x)$ into $q(z|x)$. After $q(z|x)$ is received by the decoder, it accesses the CodeBook using this index to retrieve the corresponding vector, combining them into $z_q(x)$. The decoder ultimately reconstructs the original data using $z_q(x)$. 

To train a semantic communication system based on VQVAE, we commonly use the following loss functions:
\begin{equation}
\begin{split}
L = & \log p\left(x \mid z_q(x)\right) + \left\|\operatorname{sg}\left[z_e(x)\right] - e\right\|_2^2 \\
    & + \beta\left\|z_e(x) - \operatorname{sg}[e]\right\|_2^2,
\end{split}
\end{equation}
where $sg(\cdot)$ denotes the stop-gradient operation. Since the encoder has been pre-trained in \textit{Stage 1}, the third term of the loss function could be omitted.

\section{Simulation Results}

\subsection{Implement Details}

\textbf{Architecture.} We utilize ResNet as the local encoder and the server side employs CLIP to encode the ImageNet ILSVRC-2012 dataset (ImageNet1K, which includes 1,281,167 images)~\cite{imagenet}. 

FAISS (Facebook AI Similarity Search) is a library developed by Facebook AI Research designed for efficient similarity search and clustering of dense vectors. In this paper, we adopt FAISS as vector base which stores the encoded representation vectors as the Shared Knowledge Base (\textit{SKB}). Moreover, CLIP leverages an image encoder based on ViT-B/32 and a text encoder based on transformer architecture. 

\textbf{Training.} Following the~\cite{cl_6}, we adopt Adam optimizer and use a temperature $\tau$ of 0.1. Then, we train the local encoder for 50 epochs based on CIFAR100 with learning rate of 0.005. To facilitate rapid model convergence, we implement a cosine annealing learning rate decay strategy. To match the dimensionality of the representation vectors in the \textit{SKB}, the ResNet model used by the local encoder encodes images into 512-dimensional representation vectors. During training, a batch size of 256 is used.

When retrieving from the \textit{SKB}, we add Gaussian noise with a mean of 0 and a variance of 0.2 to the representation vectors output by the query branch. We then return 30 query results and fuse them using a cross-attention mechanism

\subsection{Linear Evaluation}
Following the standard linear evaluation procedure, we add a linear classifier with 3-layers MLP on the frozen local encoder. We conduct training on CIFAR10, CIFAR100, and STL10 datasets, and evaluated the top-1 and top-5 classification accuracy respectively.

As illustrated in Table.~\ref{Table 1},~\ref{Table 2}, and~\ref{Table 3}, our proposed method shows a slight advantage over commonly used contrastive learning methods on CIFAR10 and CIFAR100. However, on STL10, our method demonstrates a more significant advantage. This is because, during the training phase, the local encoder was only train on CIFAR100 and has never been trained on STL10. Other methods produce local encoders with weaker generalization. In contrast, our local encoder, guided by the knowledge base during training, exhibits better generalization performance. This is because the shared knowledge base contains the rich representational knowledge of the ImageNet1K dataset. During the contrastive learning process, the local encoder can learn a wider variety of representations.

Additionally, we evaluate the impact of encoders of different sizes. It could be observed that encoders with a higher number of parameters slightly improved performance. Choosing the parameter size of the encoder model requires balancing the computational power of the local device.

\begin{table}[htbp]
    \centering
    \small
    \begin{tabular}{lcc}
        \toprule
        Method & \begin{tabular}[c]{@{}c@{}}Top-1(\%)\end{tabular} & \begin{tabular}[c]{@{}c@{}}Top-5(\%)\end{tabular} \\
        \midrule
        NNCLR & 84.2 & 99.2 \\
        SIMCLR & 79.3 & 99.6 \\
        BYOL & 75.6 & 98.7 \\
        MoCo & 74.3 & 98.1 \\
        SimSiam & 75.1 & 98.8 \\
        SwAV & 75.2 & 98.9 \\
        \midrule
        ours(ResNet-18) & 84.4 & 99.6 \\
        ours(ResNet-34) & 84.6 & 99.5 \\
        ours(ResNet-50) & 84.8 & 99.8 \\
        \bottomrule
    \end{tabular}
    \caption{CIFAR10 Linear Evaluation Results}
    \label{Table 1}
\end{table}

\begin{table}[htbp]
    \centering
    \small
    \begin{tabular}{lcc}
        \toprule
        Method & \begin{tabular}[c]{@{}c@{}}Top-1(\%)\end{tabular} & \begin{tabular}[c]{@{}c@{}}Top-5(\%)\end{tabular} \\
        \midrule
        NNCLR & 82.4 & 98 \\
        SIMCLR & 81.3 & 97.3 \\
        BYOL & 80.2 & 97.2 \\
        MoCo & 81.2 & 98.2 \\
        SimSiam & 80.9 & 98.7 \\
        SwAV & 80.2 & 98.6 \\
        \midrule
        ours(ResNet-18) & 83 & 98.7 \\
        ours(ResNet-34) & 84.1 & 98.5 \\
        ours(ResNet-50) & 85 & 98.7 \\
        \bottomrule
    \end{tabular}
    \caption{CIFAR100 Linear Evaluation Results}
    \label{Table 2}
\end{table}

\begin{table}[htbp]
    \centering
    \small
    \begin{tabular}{lcc}
        \toprule
        Method & \begin{tabular}[c]{@{}c@{}}Top-1(\%)\end{tabular} & \begin{tabular}[c]{@{}c@{}}Top-5(\%)\end{tabular} \\
        \midrule
        NNCLR & 84 & 99.5 \\
        SIMCLR & 83.3 & 99 \\
        BYOL & 82.2 & 99.2 \\
        MoCo & 81.2 & 98.4 \\
        SimSiam & 80.9 & 98.2 \\
        SwAV &  80.9 & 98.2 \\
        \midrule
        ours(ResNet-18) & 90.4 & 99.8 \\
        ours(ResNet-34) & 90.1 & 99.5 \\
        ours(ResNet-50) & 91.2 & 99.8 \\
        \bottomrule
    \end{tabular}
    \caption{STL10 Linear Evaluation Results}
    \label{Table 3}
\end{table}

\subsection{Data Reconstruction Evaluation}
The common paradigm of semantic communication is to use a semantic encoder to extract key semantic information, then transmit it through a channel to the receiver, where the decoder maximally reconstructs the original information. To validate the data reconstruction performance of the semantic communication system, we designed a lightweight decoder consisting of five upsampling residual blocks, with each residual block containing three layers of transposed convolution layers. 

We also froze the parameters of the local encoder, focusing solely on training the decoder. To further evaluate its performance, we utilize the \href{https://www.kaggle.com/datasets/likhon148/animal-data}{Animal} and \href{https://mmlab.ie.cuhk.edu.hk/projects/CelebA.html}{Celeba} dataset, which have not been employed during the encoder training phase, for training the decoder. Notably, unlike semantic communication research focused on high-definition image reconstruction, this work primarily aims to verify the effectiveness of the knowledge base guidance and the generalization capability of the representations provided by the local encoder.

\begin{figure}[h]
    \centering
    \noindent\includegraphics[width=3in]{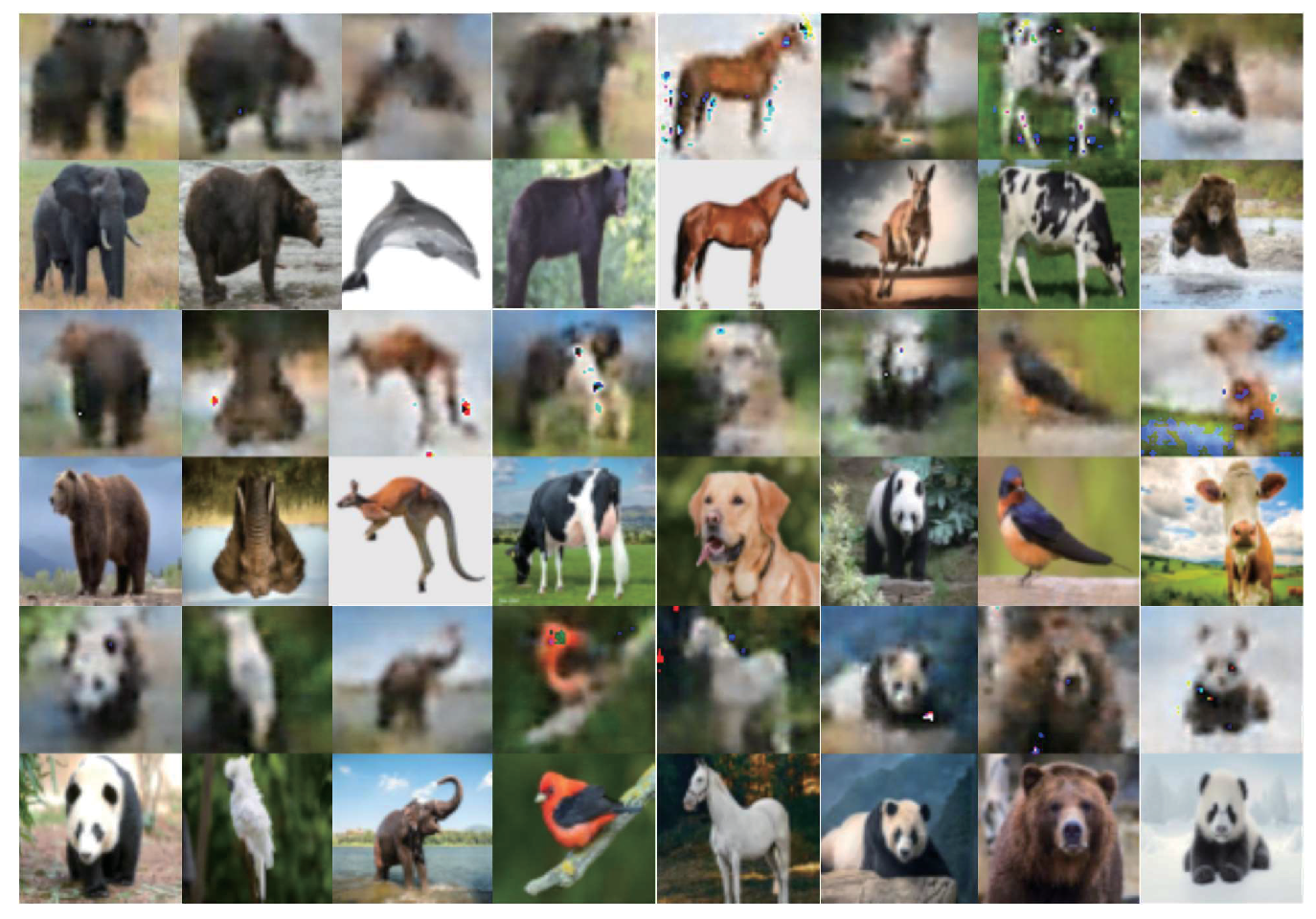}
    \caption{\textbf{Data Reconstruction on Animal Dataset}. Odd rows represent reconstructed images, while even rows represent original images.}
    \label{fig_6}
\end{figure}

\begin{figure}[h]
    \centering
    \noindent\includegraphics[width=3in]{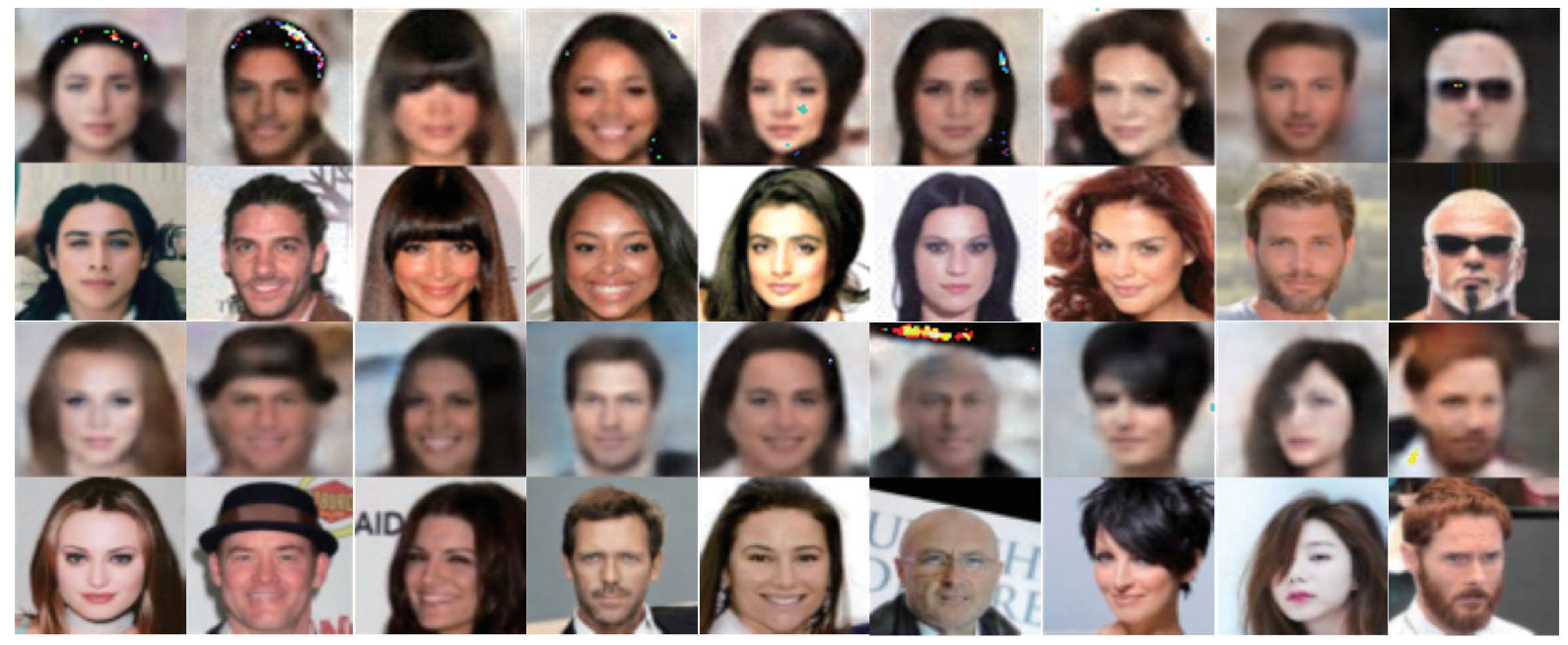}
    \caption{\textbf{Data Reconstruction on Cleba Dataset}. Odd rows represent reconstructed images, while even rows represent original images.}
    \label{fig_7}
\end{figure}

As shown in Fig.~\ref{fig_6} and~\ref{fig_7}, it can be observed that after training the decoder, the entire system is able to approximately reconstruct the information of the original image. Additionally, we subject the original images to low-light conditions to verify whether the decoder could reconstruct information under extreme conditions. 

These experimental results indicate that the encoder, trained under the guidance of a knowledge-rich knowledge base, possesses sufficient generalization ability and robustness. It can provide accurate and rich representation encodings for various downstream tasks, enabling even a simple downstream task module to perform effectively.

\subsection{Impact of \textit{SKB} Type}
To explore the impact of different types of knowledge bases on model performance, we also compare the classification performance on STL10 guided by text-based and image-based knowledge bases. Here, we uniformly used ResNet18 as the local encoder model.

\begin{figure}[htbp]
    \centering
    \noindent\includegraphics[width=2.4in]{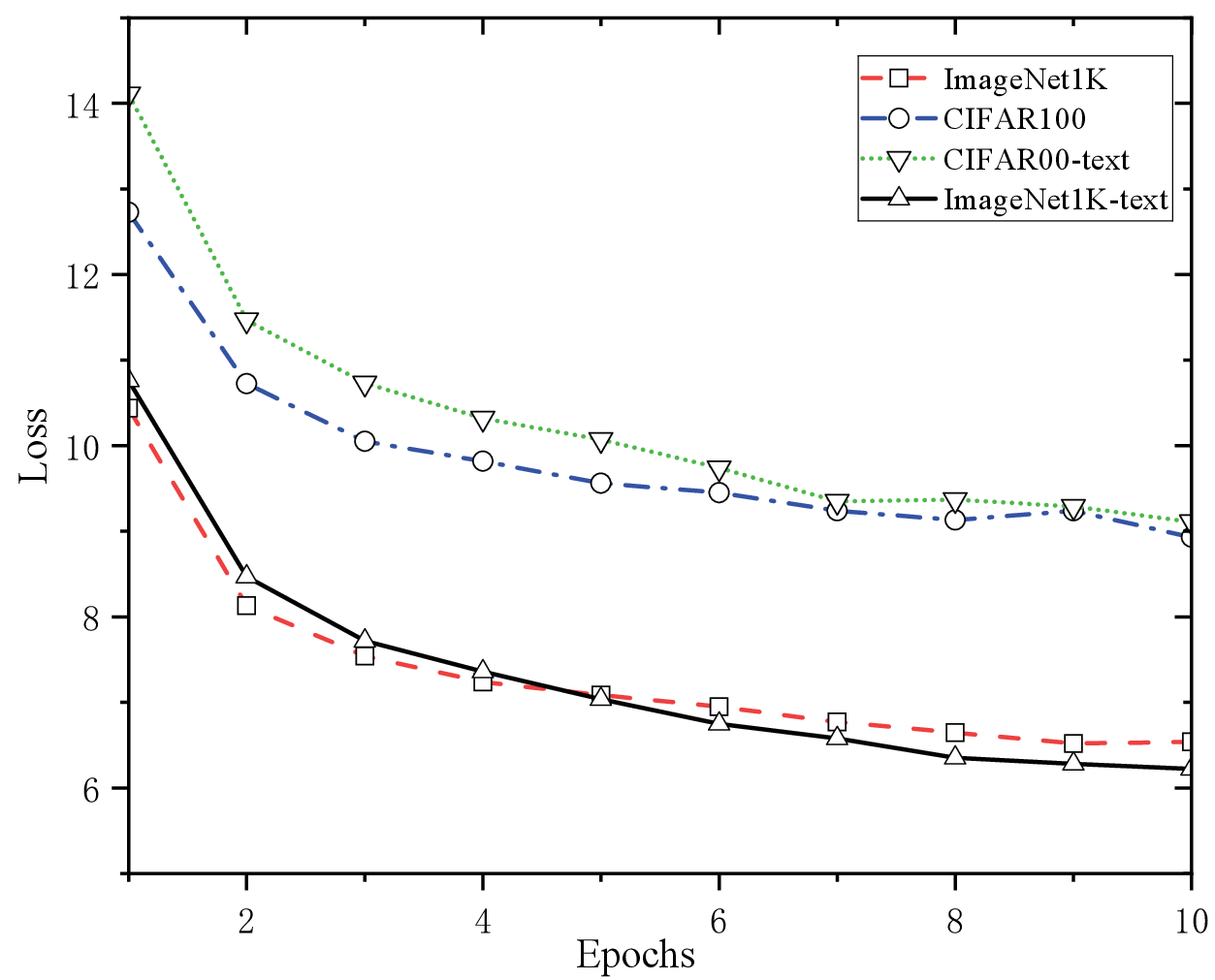}
    \caption{Loss Curve}
    \label{fig_5}
\end{figure}

As illustrated in Table.~\ref{Table_4}, it can be observed that when testing on the CIFAR series datasets, the performance of the image-based knowledge base surpasses that of the text-based knowledge base. This is because the limited category information results in less provided textual information. Interestingly, when ImageNet1K is used as the knowledge base source, the encoder trained with the guidance of the text-based knowledge base performs better in classification. This is due to the richer category information, while the large-scale image-based knowledge base contains a lot of redundant information, which can affect retrieval efficiency and accuracy during contrastive learning training.
\begin{table}
    \centering
    \small
    \begin{tabular}{lcc}
        \toprule
        \textit{SKB} & \begin{tabular}[c]{@{}c@{}}Top-1(\%)\end{tabular} & \begin{tabular}[c]{@{}c@{}}Top-5(\%)\end{tabular} \\
        \midrule
        CIFAR10 & 77.1 & 82.1 \\
        CIFAR\text{10}-text & 72.8 & 81 \\
        CIFAR100 & 81.2 & 98.1 \\
        CIFAR\text{100}-text & 73.2 & 84 \\
        ImageNet\text{1}k & 90.4 & 99.8 \\
        ImageNet\text{1}k-text & 93.4 & 99.9 \\
        \bottomrule
    \end{tabular}
    \caption{Impact of Different \textit{SKB} Type}
    \label{Table_4}
\end{table}
To further validate how text-based and image-based knowledge bases affect the training of the local encoder model, we select the loss curves for the first 10 epochs as shown in Fig.~\ref{fig_5}. It can be observed that, when training on CIFAR100, the training loss guided by the image-based knowledge base decreases faster than that guided by the text-based knowledge base. However, on ImageNet1K, we observe an interesting phenomenon: the training loss guided by the text-based knowledge base gradually decreases faster than that guided by the image-based knowledge base. This is because the ImageNet1K contains an overwhelming number of images, introducing redundant information and noise, unlike the cleaner text data.

From the above experimental analysis, we can conclude that larger knowledge bases can provide more guiding information, thereby improving model performance. However, knowledge bases with too much redundant information can also reduce model performance. Therefore, this also inspires us to refine the dataset when constructing knowledge bases.

\section{Conclusion}
This paper addresses the challenges in research on knowledge base within the field of semantic communication by proposing a simple and effective construction method. This method involves encoding large-scale general datasets using pre-trained large models and storing the representation vectors in a vector database. To facilitate the transferring of non-parametric knowledge to local encoder models, a contrastive learning-based approach is designed. This approach utilizes the shared knowledge base to provide diverse perspective samples for contrastive learning, resulting in the training of robust local encoders with strong generalization capabilities. Ultimately, by fine-tuning the decoder locally, various semantic communication tasks can be accomplished.

\newpage
\bibliographystyle{named}
\bibliography{ijcai25}

\end{document}